\documentclass[usenatbib,useAMS]{mn2e}
\usepackage{times}
\usepackage{epsfig}
\title{Dynamical response to supernova-induced gas removal in 
spiral galaxies with dark matter halo}
\author[H. Koyama, M. Nagashima, T. Kakehata, and Y. Yoshii]
{Hiroko Koyama$^{1,2}$\thanks
{E-mail: hiroko@allegro.phys.nagoya-u.ac.jp}, Masahiro Nagashima$^{3}$,
Takayuki Kakehata$^{4}$ and
Yuzuru Yoshii$^{5,6}$\\
$^{1}$Department of Physics,
Nagoya University, Chikusa-ku, Nagoya, 464-8602, Japan\\
$^{2}$Department of Physics,
Waseda University, Shinjuku-ku, Tokyo, 169-8555, Japan\\
$^{3}$Faculty of Education, Nagasaki University, Nagasaki 852-8521, Japan\\ 
$^{4}$Department of Astronomy, School of Science, The University of Tokyo, 
Bunkyo-ku, Tokyo 113-0013, Japan\\
$^{5}$Institute of Astronomy, School of Science, The University of
Tokyo, Mitaka, Tokyo 181-0015, Japan \\
$^{6}$Research Center for the Early Universe, School of Science, The
University of Tokyo, Bunkyo-ku, Tokyo 113-0033, Japan}

\begin{document}
\maketitle
\begin{abstract}
We investigate the dynamical response, in terms of disc 
size and rotation velocity, to mass loss by supernovae in the 
evolution of spiral galaxies.  A thin baryonic disc having 
the Kuzmin density profile embedded in a spherical dark 
matter halo having a density profile proposed by Navarro, 
Frenk \& White is considered.  For a purpose of comparison, 
we also consider the homogeneous and $r^{-1}$ profiles for 
dark matter in a truncated spherical halo.  Assuming for 
simplicity that the dark matter distribution is not affected 
by mass loss from discs and the change of baryonic disc matter 
distribution is homologous, we evaluate the effects of dynamical 
response in the resulting discs.  We found that the dynamical 
response only for an adiabatic approximation of mass loss can 
simultaneously account for the rotation velocity and disc size as 
observed particularly in dwarf spiral galaxies, thus reproducing 
the Tully-Fisher relation and the size versus magnitude relation 
over the full range of magnitude.  Furthermore, we found that 
the mean specific angular momentum in discs after the mass loss 
becomes larger than that before the mass loss, suggesting that 
the mass loss would occur most likely from the central disc 
region where the specific angular momentum is low.  
\end{abstract}
\begin{keywords}
  galaxies: spiral, disc -- galaxies: dwarf -- 
  galaxies: evolution -- galaxies: formation -- galaxies: haloes -- 
  large-scale structure of the universe 
\end{keywords}

\section{Introduction}
It is well-known that the luminosity $L$ of luminous spiral galaxies 
correlates tightly with their rotation velocity $V_{\rm rot}$, 
as called the Tully-Fisher relation \citep[hereafter TFR,][]{tf77}.  
The TFR is usually described as a power-law scaling relation 
$L\propto V_{\rm rot}^{\gamma}$, and the rms scatter of the TFR is 
smaller for optical passbands of longer wavelengths.  
In particular, the I-band TFR is significantly tight and its 
scatter is approximately as small as 0.4 mag \citep{w97, kff02}, 
with a power-law index $\gamma\simeq 3-3.5$ \citep[e.g.][]{pt88, s00}.  
Light at longer wavelengths from galaxies traces old late-type stars, 
unaffected by sporadic formation of young stars, and represents the 
mass of luminous matter.  Thus, the tightness of the I-band TFR 
suggests the existence of some driving mechanism that depends on mass 
of galaxies in processes of their formation.

The current standard model of cosmology is the cold dark matter (CDM) 
model.  Since the initial spectrum of density fluctuations has larger 
amplitudes for smaller scales in this model, the scenario of structure 
formation is hierarchical clustering, in the sense that smaller dark 
haloes cluster to form larger dark haloes hierarchically, thereby 
creating a scaling relation between mass $M$ and circular velocity 
$V_{\rm circ}$ of dark haloes \citep{f82, bfpr84}.
High-resolution $N$-body simulations, based on a power spectrum 
$P(k)\propto k^{n}$ with an index $n=-2.5$ appropriate for galaxy 
scales or reciprocal of wavenumber $k^{-1}$, have provided 
$M\propto V_{\rm cir}^3$, reminiscent of the TFR if a constant 
mass-to-light $M/L$ ratio is assumed \citep[e.g.][]{nfw}.
Consequently, the CDM model naturally involves a TFR-like scaling 
relation as observed for luminous spiral galaxies, because a constant 
$M/L$ is indeed known to be a good approximation to such galaxies 
\citep[e.g.][]{fg79}.

When studying the formation of galaxies originated from the growth 
of density fluctuations in the early universe, however, it is also 
necessary to take into account the effects of star formation and supernova 
(SN) explosions in individual galaxies.  In particular, dwarf galaxies 
have shallow gravitational potential wells, so that energy feedback 
from SN explosions significantly affects their evolution and hence scaling 
relations that follow.  In a framework of monolithic collapse scenario 
of galaxy formation, several authors studied scaling relations for dwarf 
galaxies in two extreme cases, i.e., the dark matter dominated case in 
which dark halo dominates the gravitational potential \citep{ds86}, 
and the baryon dominated case of self-gravitating galaxies without dark 
matter \citep{ya87}.

\citet{sn99} derived, for the first time by numerical simulations, 
the TFR for luminous spiral galaxies in the CDM model, using a method 
of $N$-body and smoothed particle hydrodynamics (SPH) combined with 
star formation and SN feedback explicitly.  They also showed that the 
star formation rate is regulated for $M/L$ to be almost constant.  
However, the zero-point of the TFR turned out to differ from that 
observed, so that the simulated spiral galaxies were too faint to be 
consistent with observations at the same circular velocity.  It was 
therefore pointed out that processes of star formation and SN feedback 
in current $N$-body/SPH simulations have to be improved 
\citep[see also][]{g07, psl07}.

Semi-analytic (SA) models of galaxy formation have been invented as 
a complementary approach to numerical simulations, in which complex 
processes such as star formation and SN feedback are simply modeled 
on galaxy scales.  SA models have indeed succeeded in reproducing 
many observational results, for example, the luminosity function of 
galaxies as well as the TFR, by suppressing the formation of dwarf 
galaxies owing to SN feedback 
\citep{sp99, c00, ny04, dlkw04, kjmb05, croton06}.

Recent progress in observational techniques has enabled to see much 
more properties of distant galaxies and faint dwarf galaxies 
beyond the limit of magnitude attained so far.  In contrast to the 
success of SA models in accounting for luminous galaxies, they 
came to face a serious discrepancy between predicted and observed 
dynamical properties of dwarf galaxies, while reproducing their 
photometric properties quite well \citep[e.g.][]{c00, vdb02}.  

Here we would like to stress that the dynamical response to gas 
removal induced by SN explosions, which has been overlooked in most 
SA models, is unavoidable in dwarf, less massive galaxies.  
\citet{ny03} formulated how the structure of spherical galaxies 
responds to SN-induced gas removal in a gravitational potential of 
dark halo, and \citet{ny04} incorporated this effect into a SA model 
to show that many properties of elliptical galaxies, including the 
Faber-Jackson relation \citep{fj76}, can be reproduced to faint 
magnitudes of dwarf ellipticals.  A more sophisticated SA model 
associated with high-resolution $N$-body cosmological simulations 
also provides a good agreement between predicted and observed properties 
of galaxies \citep{nyeyg05}.  Accordingly, as a natural extension of 
Nagashima \& Yoshii's trial, it is worth investigating whether such 
a dynamical effect could also reproduce the observed TFR down to faint 
magnitudes in spiral galaxies.

In this paper, using the Kuzmin disc \citep{k52, k56} as a  
galaxy disc embedded in a spherical dark halo, we present in 
Section 2 the general formulation for the dynamical 
response in size and rotation velocity of discs.  In Section 3 
we examine such effect for several choices of density distribution 
of dark halo.  In Section 4 we derive the resulting TFRs and disc 
size versus magnitude relations.  In Section 5 we summarize the 
results of this paper.  Detailed derivations of analytic expressions 
are given in Appendix.

\section{Dynamical Response of Galactic Disk in Dark Matter Halo}
\label{sec:formalism}
In this section, we formulate the dynamical response accompanied 
by SN feedback to a spiral galaxy consisting of baryons and dark matter. 
We consider a thin baryonic disc or the Kuzmin disk embedded in an 
non-rotating spherical dark halo whose density profile is either 
NFW, homogeneous or $r^{-1}$.  We assume that the thin disc is 
axisymmetrical and rotation-dominated with negligible velocity 
dispersions. Note that the density profile of the Kuzmin disc is 
more or less similar to the exponential disc except for the central 
region.

The gas of low angular momentum in an extended halo collapses 
towards the galaxy centre on a dynamical timescale and dissipates 
the energy to cause a burst of star formation in the central region. 
On the other hand, the gas of higher angular momentum gradually 
falls onto a disc plane on a longer timescale and settle on a 
circular annulus of the disc at the radius according to the value 
of angular momentum of the gas.  This fallen gas is converted into 
stars to form a stellar disc.

If the kinetic energy, which is released by SNe following the star 
formation, exceeds the binding energy of the disc, the remaining gas 
is removed out of the disc. This disc dynamically recovers a final 
equilibrium.  In general, the final state could be either a 
puffed-up disc where velocity dispersions dominate over the rotation, 
or a thin disc where the rotation remains to dominate with negligible 
velocity dispersions \citep{bs79}.  In this paper, it is enough to 
consider only the latter, because we are primarily interested in 
scaling relations in spiral galaxies.

First, we consider that the gas removal occurs after all the 
material falls in to form the disc.  In this case, the disc achieves 
a centrifugal equilibrium with the mass $M_i$ and the angular 
momentum $J_i$ as an initial state. Since the gas removal accompanies  
the simultaneous losses of mass and angular momentum, the disc should 
newly recover a centrifugal equilibrium with the mass $M_f$ and the 
angular momentum $J_f$ as a final state, depending not only on the 
amount of such losses $\Delta M (M_i-M_f)$ and $\Delta J (J_i-J_f)$, 
but also on whether the time scale of such losses is longer or shorter 
than the dynamical scale \citep{bs79}.

On the other hand, the gas removal may well occur while the material is 
still falling into the disc. In this case, only a part of the material 
falls in to form the disc and almost no gas removal occurs afterwards.  
The disc would then be settled in centrifugal equilibrium with the 
mass $M_f$ and the angular momentum $J_f$ throughout from the beginning. 
This situation is formally equivalent to the case such that the gas 
removal occurs on much shorter timescale compared to the dynamical 
timescale. Therefore, the related formula below in this section could 
also apply to the situation considered here. 

\subsection{Formulation}
\label{subsec:form}
The kinetic energy $T$ due to rotation, the gravitational self plus interaction 
potential energy $W$, and the total angular momentum $J$ of the Kuzmin 
disc of baryons in a dark halo are given by 
\begin{equation}
 T=\frac{1}{2}M_bV_b^2 , 
\end{equation}
\begin{equation}  
 W=-\frac{1}{4}G\frac{M_b^2}{r_b}-G\frac{M_bM_d}{r_d}f(z) , 
\end{equation}
and
\begin{equation}
\label{eq:am}
 J=gM_br_bV_b ,
\end{equation}
respectively, where $G$ is the gravitational constant and $g$ is 
the constant dependent on density distribution, $M$ is the mass, 
$V$ is the characteristic 
rotation velocity, and $r$ is the characteristic radius.  
The subscripts ``$b$'' and ``$d$'' refer to the baryonic and dark 
matter components, respectively.  The function $f(z)$ has a form 
dependent on density distribution, and the argument $z$ is a radius 
$r_b$ normalized by $r_d$ or $z=r_b/r_d$. 

Hereafter, the quantities of baryons with the subscript ``$i$'' in 
place of ``$b$'' are designated as an initial state before the gas 
removal, and those with the subscript ``$f$'' as a final state after 
the gas removal. Note that the quantities of dark matter with the 
subscript ``$d$'' are assumed not to change throughout the gas removal. 

First, we describe the initial state expressing the total energy 
of the disc as 
\begin{equation} 
\label{eq:e-initial}
 E_i=T_i+W_i=\frac{1}{2}M_iV_i^2
 -\frac{1}{4}G\frac{M_i^2}{r_i}-G\frac{M_iM_d}{r_d}f(z_i) ,
\end{equation}
provided that the virial equilibrium $2T_i=-W_i$ holds, 
and the angular momentum as 
\begin{equation}
\label{eq:ami}
 J_i=gM_ir_iV_i .
\end{equation}

Next, assuming that the disc profile does not change just after 
the gas removal, we express the total energy as  
\begin{equation}
 E_f=\frac{1}{2}M_fV_i^2
 -\frac{1}{4}G\frac{M_f^2}{r_i}-G\frac{M_fM_d}{r_d}f(z_i) .
\end{equation}
Then, conserving this energy, the disc dynamically responds 
towards recovering a virial equilibrium as the final state:
\begin{equation}
\label{eq:e-final}
 E_f=\frac{1}{2}M_fV_f^2
 -\frac{1}{4}G\frac{M_f^2}{r_f}-G\frac{M_fM_d}{r_d}f(z_f) ,
\end{equation}
and 
\begin{equation}
\label{eq:amf}
 J_f=gM_fr_fV_f .
\end{equation}

We here consider an extreme approximation of impulsive mass loss such 
that the gas removal is faster than the dynamical relaxation. 
From equations (\ref{eq:e-initial}) and (\ref{eq:e-final}), 
we express the final quantities in terms of the initial 
quantities as  
\begin{equation}
\label{eq:mfmi-in}
 \frac{m_f}{m_i}=
 \frac{1+4(z_i/m_i)[f(z_f)-f(z_i)]}{2-z_i/z_f} ,
\end{equation}
and
\begin{equation}
\label{eq:vfvi-in}
 \frac{V_f}{V_i}=
 \left[\frac{m_f/z_f+4f(z_f)}{m_i/z_i+4f(z_i)}\right]^{1/2} ,
\end{equation}
where $z_{i,f}=r_{i,f}/r_d$ and $m_{i,f}=M_{i,f}/M_d$. 

We consider another extreme approximation of adiabatic mass loss such 
that the gas removal is slower than the dynamical relaxation. 
The total change from the initial to final states is a sum of 
consecutive infinitesimal changes.  Substituting $m_f=m_i+dm$ and 
$z_f=z_i+dz$ in equation (\ref{eq:mfmi-in}) and linearizing it, we 
obtain 
\begin{equation}
 \frac{dm}{dz}=-\frac{m}{z}+4z\frac{df(z)}{dz} ,
\end{equation}
where the subscript ``$i$'' is omitted for simplicity. The first term 
comes from the self-gravitating baryons alone, and the second term 
from the gravitational interaction of baryons with dark halo.  Solving 
this differential equation, we obtain 
\begin{equation}
 m=\frac{C}{z}+q(z) ,
\end{equation}
where 
\begin{equation}
 q(z)\equiv \frac{b}{az}\int_0^{z}t^2\frac{df(t)}{dt}dt ,
\label{eq:q}
\end{equation}
and $C$ is an integration constant or an adiabatic invariant here. 
Then we express the final quantities in terms of the initial 
quantities for the adiabatic mass loss:
\begin{equation}
\label{eq:mfmi-ad}
 \frac{m_f}{m_i}=\frac{C/z_f+q(z_f)}{m_i}
 =\frac{z_i}{z_f}+\frac{q(z_f)-z_iq(z_i)/z_f}{m_i} , 
\end{equation}
together with the same form of equation (\ref{eq:vfvi-in}) for the 
rotation velocity. 

It is apparent from equations (\ref{eq:mfmi-in}), (\ref{eq:vfvi-in}),
and (\ref{eq:mfmi-ad}) that the ratios of dynamical quantities
$r_f/r_i(=z_f)$ and $V_f/V_i$ are obtained when the ratios of
$m_i(=M_i/M_d)$, $z_i(=r_i/r_d)$, and $m_f/m_i(=M_f/M_i)$ are given. In
other words, given $m_i$ and $z_i$, the ratios of $r_f/r_i$ and
$V_f/V_i$ are formally written as a function of $M_f/M_i$:
\begin{equation}
\label{rel:rfri}
 \frac{r_f}{r_i}=H(m_i,z_i; M_f/M_i) ,
\end{equation}
and
\begin{equation}
\label{rel:vfvi}
 \frac{V_f}{V_i}=I(m_i,z_i; M_f/M_i) .
\end{equation}

For the impulsive and adiabatic approximations of mass loss, 
the change of specific angular momentum is given by
\begin{equation}
\label{eq:dam}
 \frac{(J/M)_f}{(J/M)_i}=\frac{r_fV_f}{r_iV_i} ,
\end{equation}
from equations (\ref{eq:ami}) and (\ref{eq:amf}).  When the gas removal 
occurs from the central region of the disc where the mass is 
concentrated but the rotation velocity is small, $(J/M)_f/(J/M)_i$ 
should exceed unity.  On the other hand, when the gas removal 
occurs from the outer region of the disc where the mass is small 
but the rotation velocity is large, $(J/M)_f/(J/M)_i$ is less than 
unity.  Therefore we use the value of $(J/M)_f/(J/M)_i$ to discuss 
where in the disc the site of efficient gas removal is located.  

\subsection{Limiting cases}
\label{subsec:limit}
In this subsection, before studying more general gas removal, 
we examine two limiting cases of $M_i/M_d\gg 1$ and $M_i/M_d\ll 1$.

In the baryon dominated case of $M_i/M_d\gg 1$, we have for the 
impulsive mass loss 
\begin{eqnarray}
 \frac{r_f}{r_i}&=&\frac{M_f/M_i}{2M_f/M_i-1} , \nonumber\\
 \frac{V_f}{V_i}&=&\sqrt{2\frac{M_f}{M_i}-1} , \nonumber\\
 \frac{(J/M)_f}{(J/M)_i}&=&
  \left(\frac{M_f/M_i}{2M_f/M_i-1}\right)
  \sqrt{2\frac{M_f}{M_i}-1}>1 ,
\end{eqnarray}
and for the adiabatic mass loss 
\begin{eqnarray}
\label{eq:limit_ba}
 \frac{r_f}{r_i}&=&\frac{M_i}{M_f} , \nonumber\\
 \frac{V_f}{V_i}&=&\frac{M_f}{M_i} , \nonumber\\
 \frac{(J/M)_f}{(J/M)_i}&=&1 .
\end{eqnarray}
The fact that the value of $(J/M)_f/(J/M)_i$ is always larger than 
unity for the impulsive mass loss suggests that the gas removal 
should preferentially occurs from the central region of the disc, 
whereas the fact of $(J/M)_f/(J/M)_i=1$ for the adiabatic mass loss 
yields that the same fraction of mass is lost across the disc over 
its entire radial range. 

In the dark matter dominated case of $M_b/M_d\ll 1$, the disc size 
and rotation velocity of the initial disc hardly change during the 
gas removal, because the dominant gravitational potential 
of dark halo allows almost no response in structure and dynamics 
of the disc. 

\section{Dynamical response on disk size and rotation velocity}
\label{sec:response}
In this section, based on the analytic expression of $f(z)$ 
for the Kuzmin disc embedded in a dark halo with the NFW, 
homogeneous, and $1/r$ density distributions, we calculate the 
strength of dynamical response for $r_f/r_i$, $V_f/V_i$, and 
$(J/M)_f/(J/M)_i$ as a function of $M_f/M_i$ for several 
combinations of $m_i=0.05-0.2$ and $z_i=0.05-0.2$. 

The range of $m_i$ explored is based on the following 
consideration. We assume that all the baryonic matter is 
distributed in the same way as the dark matter initially 
and falls in to make up the disc. Then the upper bound of 
$m_i$ is set by the current estimate of density ratio 
$\rho_b/\rho_d \simeq 0.2$ in the universe. Considering 
a possibility that not all of the baryonic matter falls into 
the initial disc, we rather arbitrarily adopt a range of 
$m_i=0.05-0.2$ for the initial mass ratio. 

The range of $z_i=r_i/r_d$ is based on the following consideration.  
From the observed luminosity profile and rotation curve of dwarf 
spirals, the scale radius of dark halo of $10^{9-10}M_{\odot}$ 
is estimated as $(1.5-4)h^{-1}$kpc, and the corresponding rotation 
velocity is estimated as $40-70\,$km$\,$s$^{-1}$ \citep{bu95}, 
where $h$ is the Hubble constant $H_0$ defined as 
$h=H_0/100\,$km$\,$s$^{-1}$Mpc$^{-1}$.  In addition, the 
radial exponential scalelength of such dwarf spirals is estimated 
as $(0.5-2)h^{-1}$kpc \citep{pse,s97}, thus giving the scale ratio of 
$0.3-0.5$ as observed in the local universe.  Regarding this ratio 
as obtained after the initial disc expands by a factor of several 
due to the gas removal (see below), we adopt a range of $z_i=0.05-0.2$ 
for the initial scale ratio.

\subsection{Kuzmin disc in the NFW halo}
\label{subsec:kuzmin-nfw}
We examine the dynamical response to the Kuzmin disc embedded 
in the NFW halo. The surface mass density distribution of the 
Kuzmin disc is given by 
\begin{eqnarray}
\label{eq:density-kuz}
 \Sigma(r)&=&\frac{1}{4\pi G}
 \left\{\left(\frac{\delta \Phi}{\delta z}\right)_{z\to 0+0}
 -\left(\frac{\delta \Phi}{\delta z}\right)_{z\to 0-0}\right\}
 \nonumber\\
 &=&\frac{M_br_b}{2\pi(r^2+r_b^2)^{3/2}} ,
\end{eqnarray}
with the gravitational potential
\begin{equation}
\label{eq:pot-kuz}
\Phi(r,z)=-\frac{GM_b}{\sqrt{r^2+(r_b+|z|)^2}} .
\end{equation}
As for the NFW halo, the density distribution is given by 
\begin{equation}
\label{eq:density-nfw}
 \rho(r)=\rho_d c^3
   \left[\frac{cr}{r_d}\left(1+\frac{cr}{r_d}\right)^2\right]^{-1} ,
\end{equation}
with the gravitational potential 
\begin{equation}
 \Phi(r)=-4\pi G\frac{\rho_dr_d^3}{r}\ln\left(1+\frac{cr}{r_d}\right) ,
\end{equation}
where $c$ is a concentration parameter. Since the essential 
characteristic length of the NFW halo is $r_d/c$ instead of $r_d$, 
we hereafter define $z=r_b/(r_d/c)$.

\begin{figure*}
   \epsfig{file=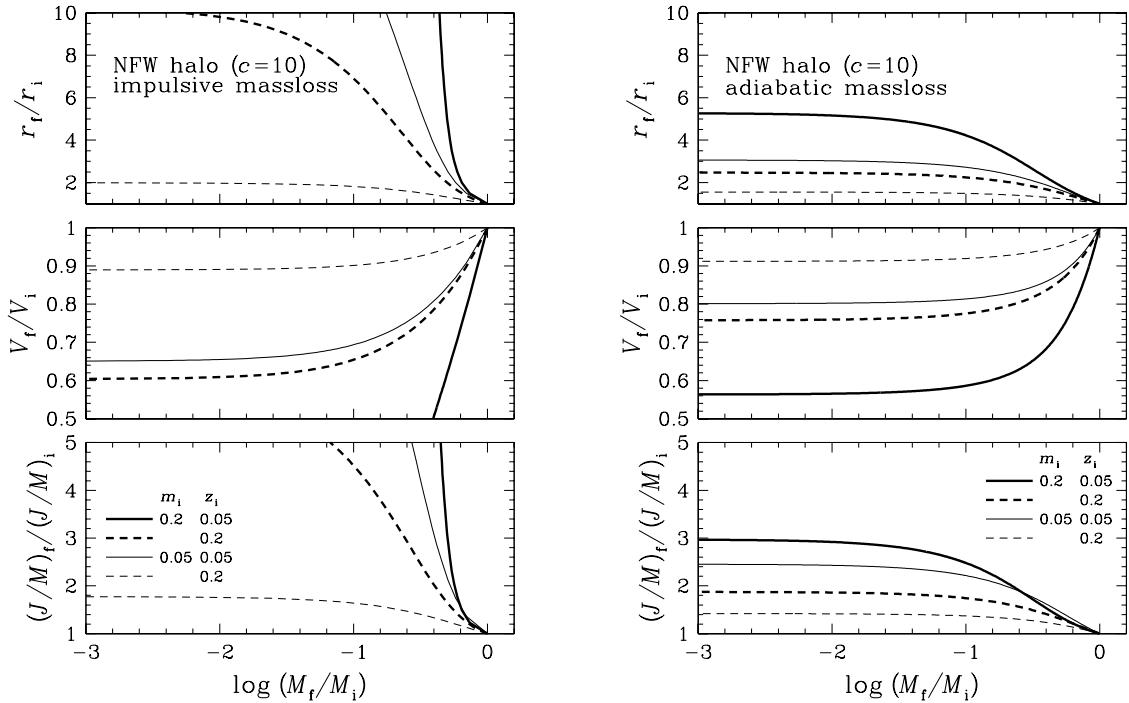,width=0.9\hsize}
 \caption{Dynamical response of the Kuzmin disc embedded in the 
 NFW halo. The results shown are the 
 dependence of dynamical response on $m_i=M_i/M_d$ and 
 $z_i=r_i/(r_d/c)$ with $c=10$ for the impulsive mass loss in the 
 left column, and for the adiabatic mass loss in the right column.  
 In each of the columns, the strength of dynamical response for 
 $r_f/r_i$ (top panel), $V_f/V_i$ (middle panel), and 
 $(J/M)_f/(J/M)_i$ (bottom panel) is plotted against the final 
 to initial disc mass ratio $M_f/M_i$ for various combinations of 
 $m_i$ and $z_i$.
}
\label{fig:res-nfw-mz}
\end{figure*}

Figure \ref{fig:res-nfw-mz} shows the dependence of dynamical 
response on $m_i$ and $z_i$ with $c=10$ for the impulsive mass loss 
in the left column, and for the adiabatic mass loss in the right 
column.  In each of the columns, the strength of dynamical 
response for $r_f/r_i$, $V_f/V_i$, and $(J/M)_f/(J/M)_i$ is shown 
in order from top to bottom panels as a function of $M_f/M_i$ for 
several combinations of $m_i$ and $z_i$.

It is clear from this figure that the dynamical response monotonically 
increases as $M_f/M_i$ decreases and asymptotically reaches a constant 
level at $M_f/M_i\le 0.1$.  Furthermore, given $M_f/M_i$, the dynamical 
response is stronger for increasing the ratio of surface mass densities 
of initial disc and dark halo projected onto the disc plane or 
$m_i/z_i^2$. Such a tendency holds irrespective of either the impulsive 
or adiabatic mass loss. However, comparing the left and right columns, 
we find that the dynamical response is greatly suppressed for the 
adiabatic mass loss, in particular, even by a factor of ten or so 
within the range of $m_i/z_i^2$ and $M_i/M_d$ considered. 

\begin{figure*}
   \epsfig{file=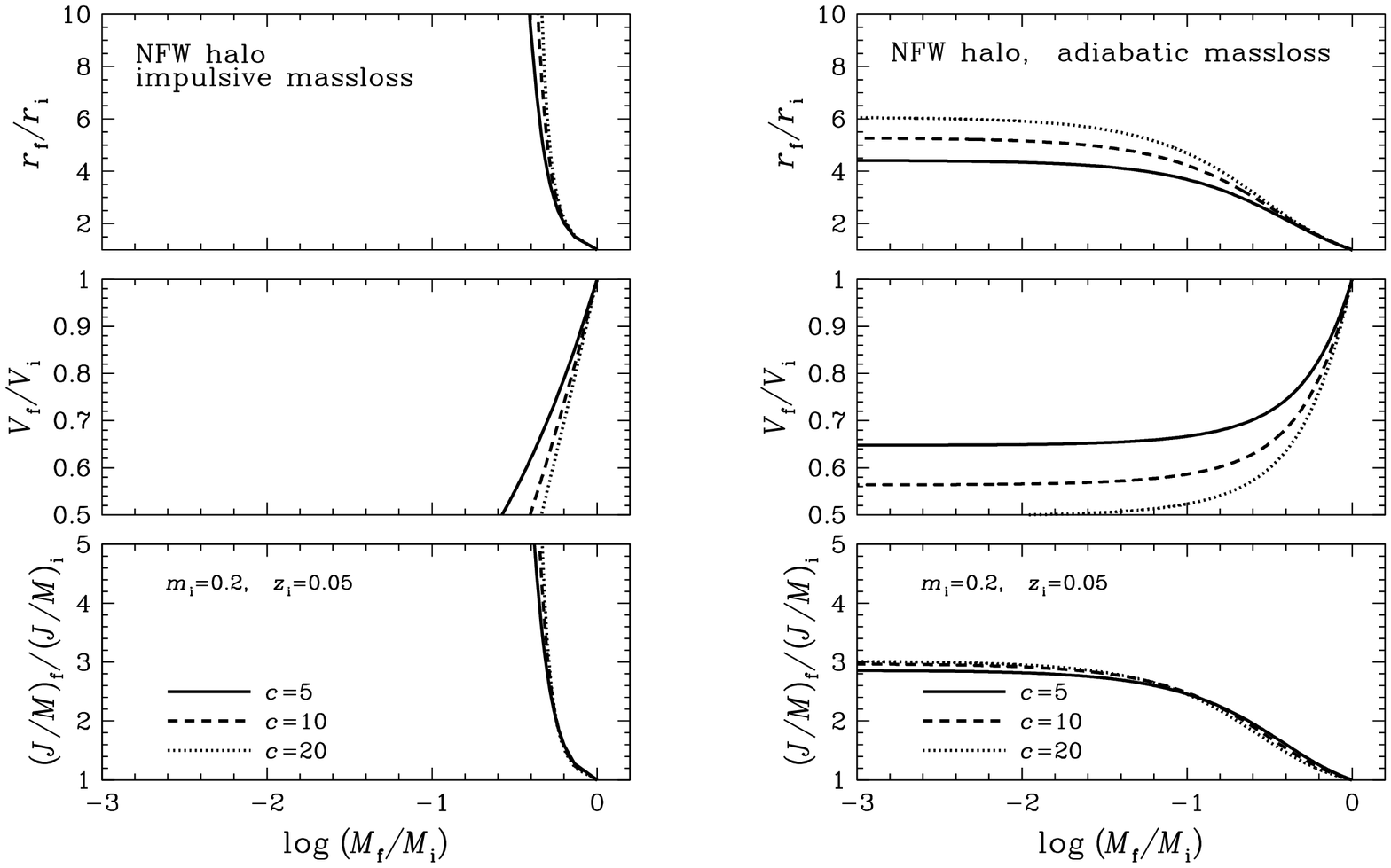,width=0.9\hsize}
 \caption{Dynamical response of the Kuzmin disc embedded in the 
 NFW halo.  Same as in Figure \ref{fig:res-nfw-mz}, but for the 
 dependence of dynamical response on the concentration parameter 
 $c$ with $m_i=0.2$ and $z_i=r_i/(r_d/c)=0.05$.  
}
\label{fig:res-nfw-c}
\end{figure*}

Figure \ref{fig:res-nfw-c} shows the dependence of dynamical 
response on $c$ with $m_i=0.2$ and $z_i=0.05$. There is a tendency 
such that the dynamical response is stronger for increasing $c$, or 
equivalently decreasing $r_i/r_d$ because $z_i=r_i/(r_d/c)$ is 
fixed here.   

The specific angular momentum after the gas removal $(J/M)_f$ is 
always larger than that before the gas removal $(J/M)_i$, which 
suggests that the gas removal occurs from the central region of 
the disc where specific angular momentum is small.  This is the 
condition that the disc should remain thin and rotation dominated 
before and after the gas removal.

\subsection{Dependence on dark matter distribution}
We examine how the Kuzmin disc responds depending on the density 
distribution of dark halo. Here we consider the homogeneous 
and $1/r$ density distributions in a spherical halo truncated at 
$r=r_d$: 
\begin{equation}
 \rho(r)=\rho_d\theta(r_d-r) ,
\end{equation}
and 
\begin{eqnarray}
 \rho(r)=\rho_d \frac{r_d}{r}\theta(r_d-r) ,
\end{eqnarray}
respectively, where $\theta(x)$ is the Heaviside step function. 

\begin{figure*}
   \epsfig{file=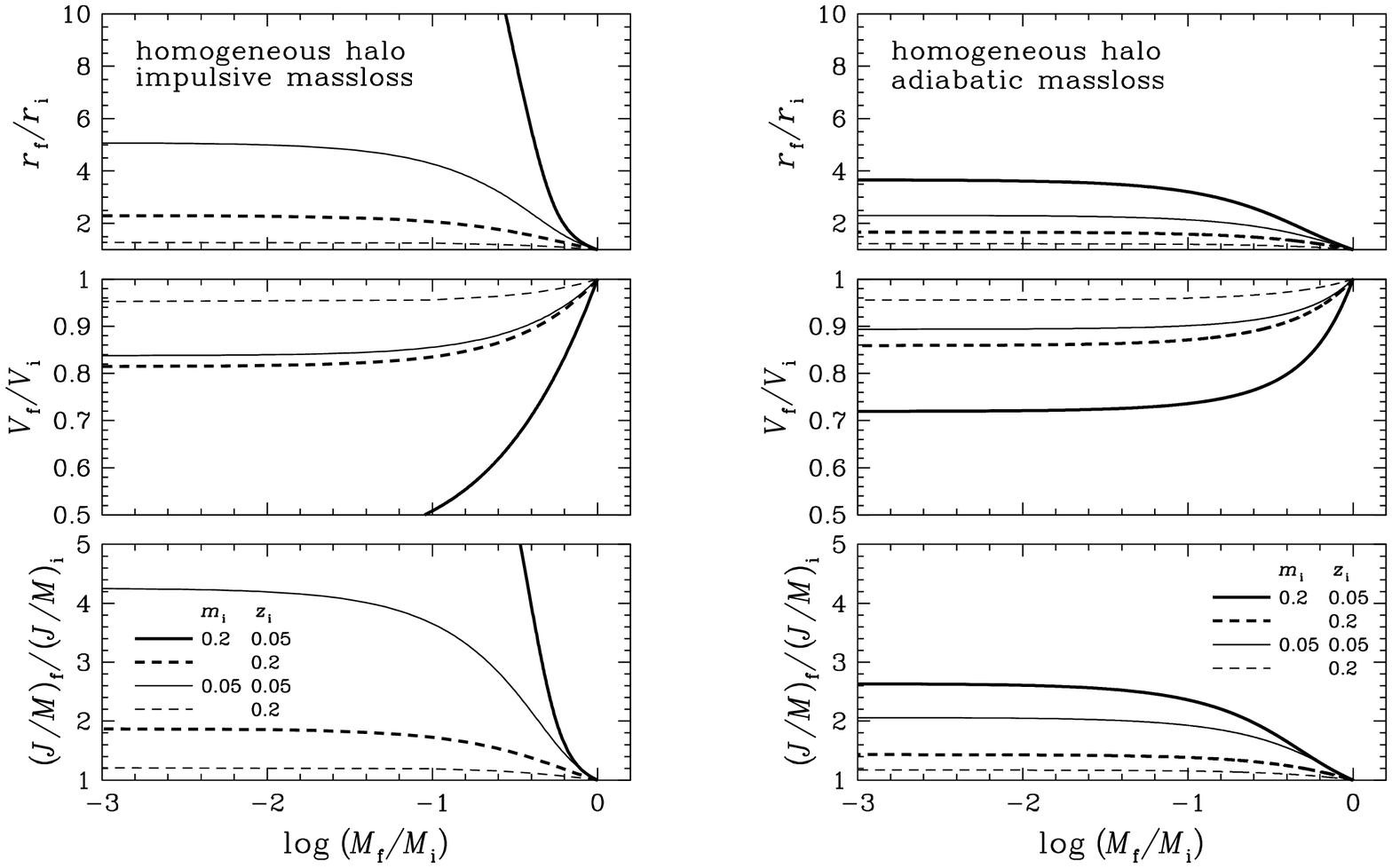,width=0.9\hsize}
 \caption{Dynamical response of the Kuzmin disc embedded in the 
 dark halo.  Same as Figure \ref{fig:res-nfw-mz}, but for the  
 truncated spherical halo with the homogeneous density distribution. 
}
\label{fig:res-homo}
\end{figure*}

\begin{figure*}
   \epsfig{file=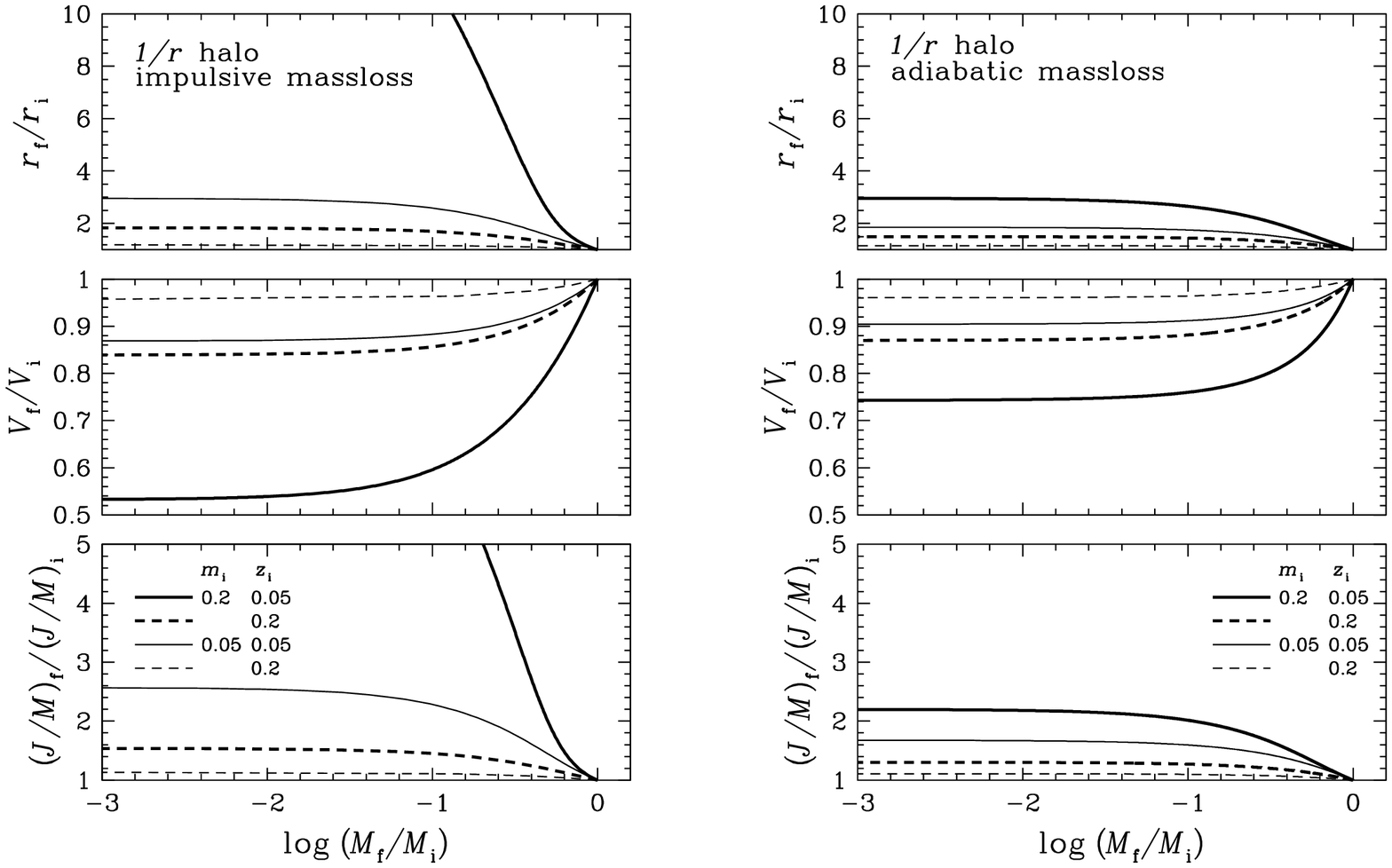,width=0.9\hsize}
 \caption{Dynamical response of the Kuzmin disc embedded in the 
 dark halo.  Same as Figure \ref{fig:res-nfw-mz}, but for the 
 truncated spherical halo with the $1/r$ density distribution. 
}
\label{fig:res-1r}
\end{figure*}

The results for these two distributions are shown in Figures 
\ref{fig:res-homo} and \ref{fig:res-1r}, respectively.  We see 
from these figures that, given $m_i$ and $z_i$, the dynamical 
response for the homogeneous density distribution is stronger 
than that for the $1/r$ density distribution with which dark 
matter is more confined inside the characteristic radius of 
the initial disc $r_i$.

We also see that, given $m_i$ and $z_i$, the dynamical response 
for the NFW halo (Figure \ref{fig:res-nfw-mz}) is even stronger 
than those considered here.  This is because, adopting $c=10$ in 
the definition of $z_i=r_i/(r_d/c)$, the NFW halo is more 
extended beyond the truncation radius $r_d$, thereby dark matter 
is less confined inside the initial disc radius $r_i$ as far as 
the value of $m_i$ is fixed.  

Furthermore, the dynamical response for the impulsive mass loss is 
much stronger than that for the adiabatic mass loss, which is also 
the case of the NFW halo (Figures \ref{fig:res-nfw-mz} and 
\ref{fig:res-nfw-c}). 

\subsection{Dependence on baryonic matter distribution}
It is instructive here to additionally examine the dynamical 
response of an exponential disc for which the surface mass density 
distribution is given by  
\begin{equation}
\label{eq:density-exp}
 \Sigma(r)=\frac{M_b}{2\pi r_s^2}\exp(-r/r_s) ,
\end{equation}
where $r_s$ is the scale length of the disc.  This choice of 
$\Sigma(r)$, which allows no analytic expression of $W$, 
is known to be more appropriate to spiral galaxies than the 
Kuzmin disc.

\begin{figure*}
   \epsfig{file=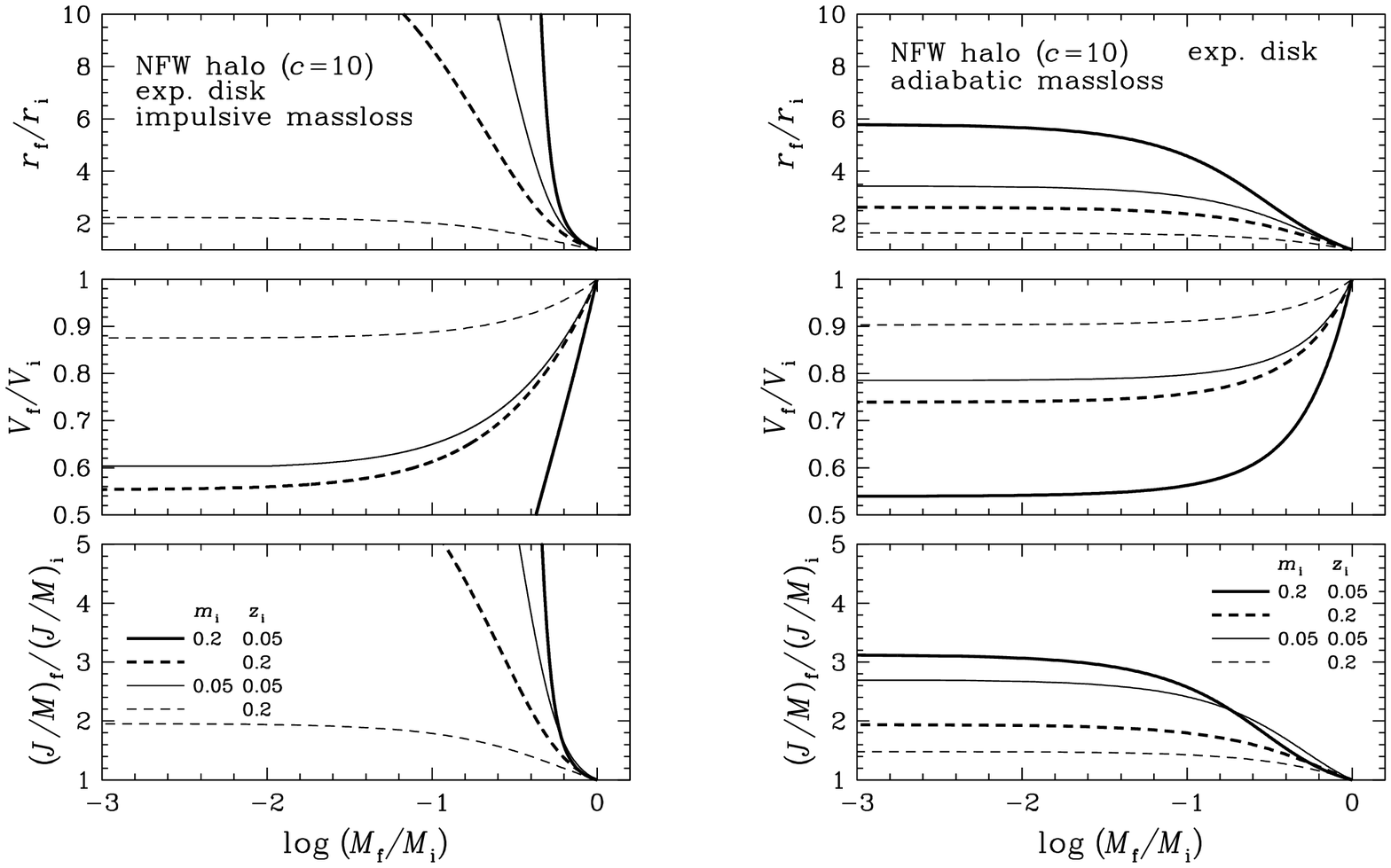,width=0.9\hsize}
 \caption{
Dynamical response of the exponential disc embedded in 
the NFW halo.  The results are shown in the same way as 
in Figure \ref{fig:res-nfw-mz}, for direct comparison with 
the results for the Kuzmin disc.  We note that the scale 
length $r_s$ of the exponential disc is related to $r_b$ in 
$z$ via $r_s=(\sqrt{3}/1.68)r_b$.       
}
\label{fig:exp}
\end{figure*}

For the purpose of comparison, we equate the effective radius 
of the exponential disc to that of the Kuzmin disc, or 
$r_e=1.68r_s=\sqrt{3}r_b$. Then, the dynamical response of 
the exponential disc embedded in the NFW halo is calculated 
by numerically integrating $W$, and the result is shown in 
Figure \ref{fig:exp} in the same way as in 
Figure \ref{fig:res-nfw-mz}.  

It is evident from these figures that, given $m_i$ and $z_i$, 
the dynamical response of the exponential disc is only slightly 
stronger than that of the Kuzmin disc. Namely, the difference 
is very small for the impulsive mass loss and is much smaller 
for the adiabatic mass loss.  This is understood, because the 
Kuzmin disc is more or less similar to the exponential disc 
except for the central region.  Therefore, without loss of 
generality, we use the Kuzmin disc in Section 4 for the study 
of dynamical response in spiral galaxies over a full range of 
magnitude observed. 

\section{Scaling relations modified by dynamical response}
\label{sec:tfr}
In this section, using the Kuzmin disc, we investigate the effects 
of dynamical response in predicting the rotation velocity versus 
magnitude relation (TFR) and the disc size versus magnitude relation. 

We begin with the assumption that dark matter haloes in the CDM 
universe obey the scaling relations \citep[e.g.][]{nfw}:
\begin{equation}
\label{eq:scaling}
 M_d\propto V_{circ}^{\gamma}\propto r_d^{\epsilon} .
\end{equation}
The baryonic matter in an individual halo falls in to form the 
galaxy disc and a part of such fallen gas goes back to the halo 
region owing to the gas removal induced by SNe in the disc.  
Cosmological $N$-body/SPH simulations have shown that along this 
scheme of galaxy formation the above scaling relations of dark 
matter haloes almost hold in bright spiral galaxies where the 
effect of SN feedback is minimal \citep[e.g.][]{sn99}. 
Consequently, in this paper, we assume that initial discs before 
the gas removal obey the same scaling relations as dark matter 
haloes, instead of entering into detailed discussions of physical 
processes involved.  This assumption accords with our use of 
$m_i=M_i/M_d$ and $z_i=r_i/r_d$ as constant parameters, yielding 
\begin{equation}
\label{eq:scalingdsk}
 M_i\propto V_i^{\gamma}\propto r_i^{\epsilon} ,
\end{equation}
where the subscript ``$i$'' stands for the quantities of initial 
discs before the gas removal as in previous section. 

Next, we express the quantities of final discs after the gas 
removal, to which the subscript ``$f$'' is attached.  
Considering that disc mass is a sum of remaining gas and stars 
$M_i=M_{gas}+M_{\ast}$ before the gas removal and is left with stars 
$M_f=M_{\ast}$ after the gas removal, and introducing the strength 
parameter for SN feedback $\beta\equiv M_{gas}/M_{\ast}$ for 
which $\beta \ll 1$ for massive and normal galaxies and 
$\beta \gg 1$ for dwarf galaxies, we obtain 
\begin{equation}
\label{eq:mimf-beta}
 \frac{M_f}{M_i}=\frac{1}{1+\beta}.
\end{equation}
Following the definition in equation (10) of \citet{nyeyg05}, 
we set 
\begin{equation}
\label{eq:beta}
 \beta=\left(\frac{V_i}{V_{hot}}\right)^{-\alpha_{hot}}  ,
\end{equation}
where $V_{hot}$ is an effective SN-feedback strength in units 
of velocity, and $\alpha_{\rm hot}$ is a constant power index.

Our analysis below confines to the NFW halo, because the 
homogeneous and $1/r$ density distributions are of academic 
interest only.  Values of the parameters used are 
$V_{hot}=150\,$km$\,$s$^{-1}$ and $\alpha_{hot}=2$ and 4 for the SN-feedback 
strength, while $m_i=0.2$ and $z_i=0.05$, 0.1, and 0.2 for the 
initial disc.

\subsection{Tully-Fisher relation}
Theoretical TFR or the $V_f-M_f$ relation is obtained by 
a set of the following equations: 
\begin{equation}
\label{eq:sn-mimf}
 M_f\propto \frac{V_i^{\gamma}}{1+\beta} ,
\end{equation}
and
\begin{equation}
 V_f=\left\{ 
\begin{array}{ll}
V_i  &  {\rm (no~ response)}  \nonumber\\  
V_i I[m_i,z_i; 1/(1+\beta)] &
    {\rm (with~ response)} ,
\end{array}\right.
\end{equation}
where $I(a,b;x)$ is given in equation (\ref{rel:vfvi}).  

\begin{figure*}
   \epsfig{file=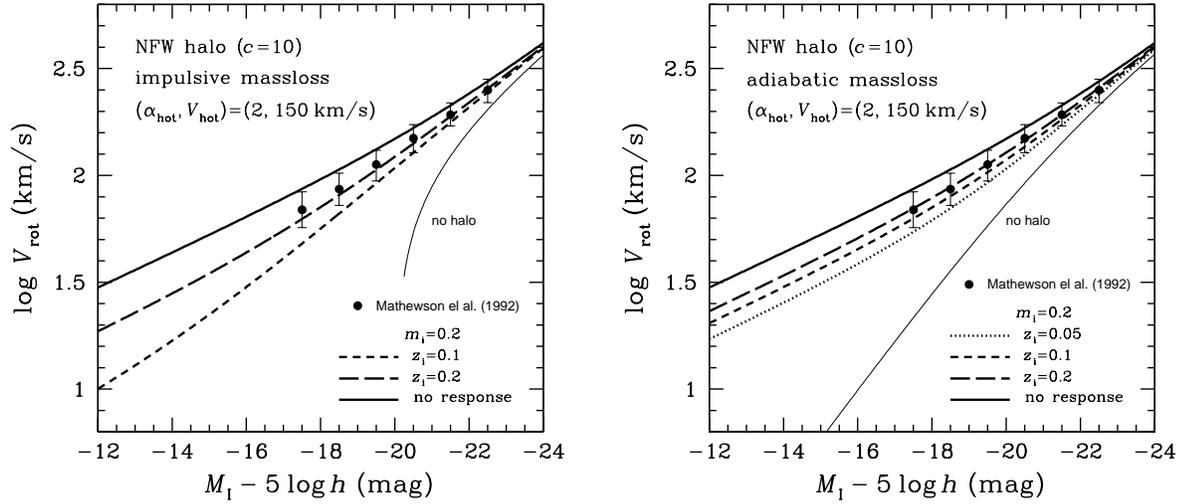,width=0.9\hsize}  
\caption{Theoretical I-band Tully-Fisher relations with and 
 without dynamical response of the Kuzmin disc embedded in the 
 NFW halo. The results with $\alpha_{hot}=2$ are shown by various 
 thick lines in the left panel for the impulsive mass loss and in 
 the right panel for the adiabatic mass loss, while the result 
 with no halo is shown by thin line only for the academic purpose 
 of comparison. Filled circles represent the data taken from the 
 table of \citet{mfb92}. $V_{rot}$ is the rotation velocity, 
 $M_{\rm I}$ is the absolute I-magnitude, and $h$ is the Hubble 
 constant $H_0$ defined as $h=H_0/100\,$km$\,$s$^{-1}$Mpc$^{-1}$.
}
\label{fig:tfr-nfw-alpha2}
\end{figure*}

\begin{figure*}
   \epsfig{file=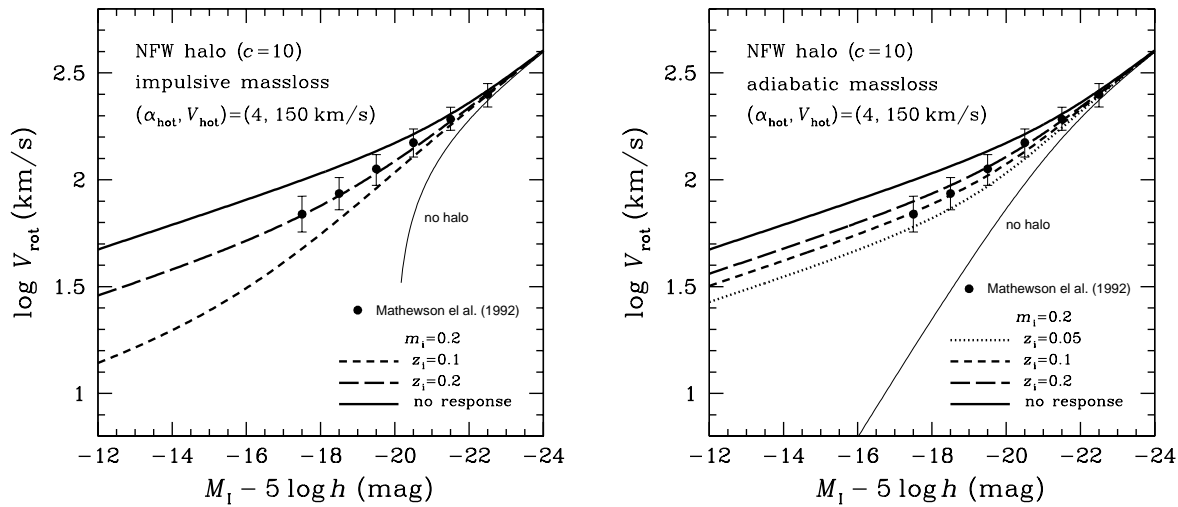,width=0.9\hsize}
\caption{Theoretical I-band Tully-Fisher relations with and 
 without dynamical response of the Kuzmin disc embedded in the 
 NFW halo. Same as in Figure \ref{fig:tfr-nfw-alpha2}, but for 
 $\alpha_{hot}=4$.
}
\label{fig:tfr-nfw-alpha4}
\end{figure*}

This relation is compared with the data of rotation velocity 
$V_{rot}$ and absolute magnitude $M_{\rm I}$ of spiral galaxies 
in the I band taken from the table of \citet{mfb92}, and the results 
are shown in Figure \ref{fig:tfr-nfw-alpha2} for $\alpha_{hot}=2$ 
and in Figure \ref{fig:tfr-nfw-alpha4} for $\alpha_{hot}=4$, 
together with the data. In each of these figures the left and right 
panels are for the impulsive and adiabatic mass losses, respectively. 

Here, assuming a constant baryonic mass to I-band light ratio 
$M_i/L_{\rm I}$ and equating $V_i$ to the observed rotation velocity 
$V_{rot}$ for bright galaxies, we have set the power index 
$\gamma=3$ in equation (\ref{eq:sn-mimf}) by adjusting the 
$V_f-M_f$ relation along the horizontal axis in Figures 
\ref{fig:tfr-nfw-alpha2} and \ref{fig:tfr-nfw-alpha4} to fit 
to the observed TFR in the bright end. Our setting of $\gamma=3$ 
is consistent with the scaling relation of dark haloes obtained 
by $N$-body CDM simulations \citep{nfw} as well as the scaling 
relation of spiral galaxies obtained by $N$-body/SPH CDM 
simulations \citep{sn99}.

We see from these figures that the $V_f-M_f$ relations with no 
dynamical response and with no dark halo deviate significantly 
from the faint data, while the dynamical response with dark halo 
improves the fit to the data. In particular, the $V_f-M_f$ 
relations that well agree with the data over the full range of 
$M_{\rm I}$ observed are those of $z_i=0.2$ and $\alpha_{hot}=2-4$ 
for the impulsive mass loss, and those of $z_i=0.1-0.2$ and 
$\alpha_{hot}=2$ as well as $z_i=0.2$ and $\alpha_{hot}=4$ for 
the adiabatic mass loss.


\subsection{Disk size versus magnitude relation}
Theoretical disc size versus magnitude relation or the $r_f-M_f$ 
relation is obtained by a set of the following equations: 
\begin{equation}
\label{eq:sn-mimf2}
 M_f\propto \frac{r_i^{\epsilon}}{1+\beta} ,
\end{equation}
and 
\begin{equation}
 r_f=\left\{
\begin{array}{ll}
r_i &  {\rm (no~ response)}\\  
r_i H[m_i,z_i;1/(1+\beta)] & {\rm (with~ response)} ,
\end{array}
\right.
\end{equation}
where $H(a,b;x)$ is given in equation (\ref{rel:rfri}). Here, $V_i(r_i)$
is obtained from the relation of
$(V_i/V_{hot})^{\gamma}=(r_i/r_{hot})^{\epsilon}$ according to equation
(\ref{eq:scalingdsk}), where $r_{hot}$ corresponding to $V_{hot}$ is
obtained from the rotation velocity versus magnitude relation and the
disc size versus magnitude relation. So far we have used the
characteristic radius $r_b$ to represent the disc size. However, we
hereafter use the effective radius $r_e=\sqrt{3}r_b$ of the Kuzmin disc
to be consistent with the usual observational definition.

\begin{figure*}
   \epsfig{file=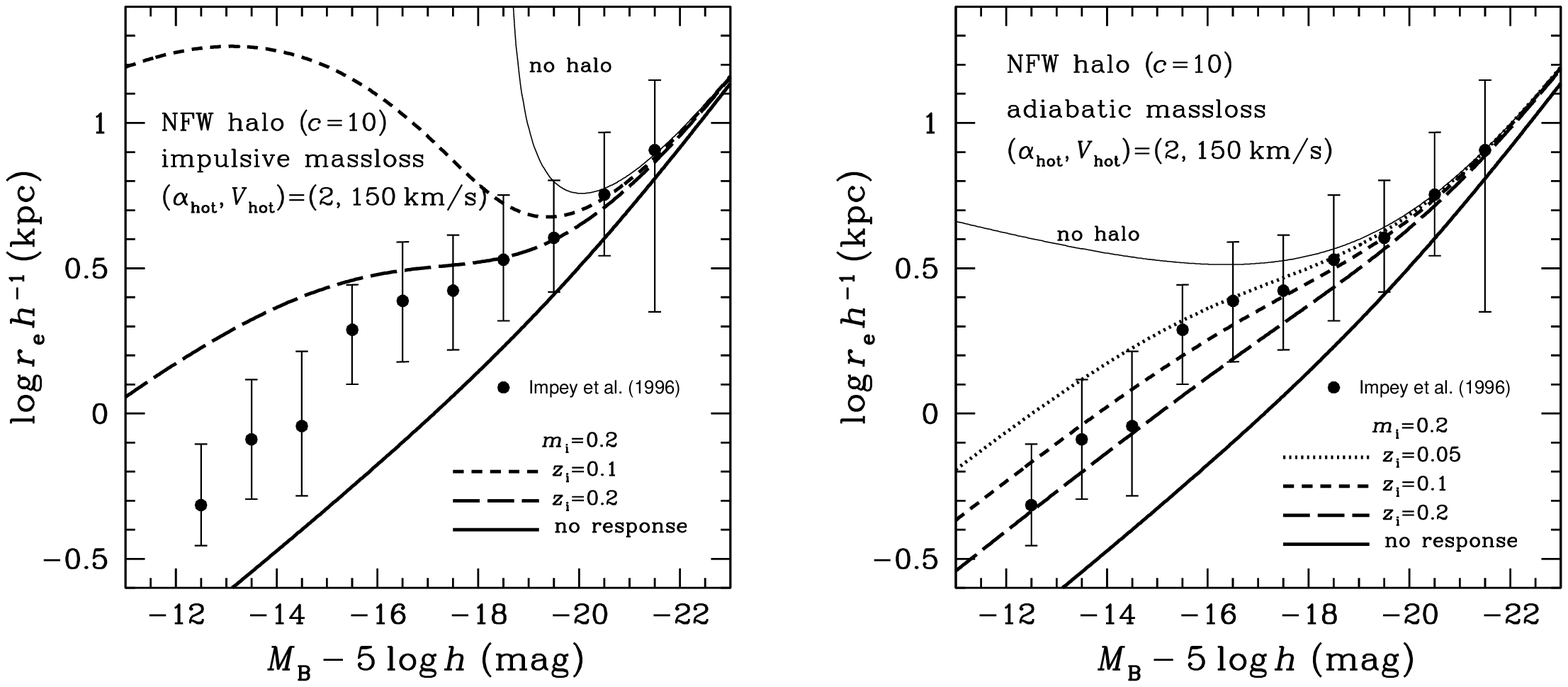,width=0.9\hsize}
 \caption{Theoretical B-band disc size versus absolute magnitude 
 relations with and without dynamical response of the Kuzmin disc 
 embedded in the NFW halo. The results with $\alpha_{hot}=2$ are 
 shown by various thick lines in the left panel for the impulsive 
 mass loss and in the right panel for the adiabatic mass loss, while 
 the result with no halo is shown by thin line only for the academic 
 purpose of comparison. Filled circles represent the data taken 
 from the table of \citet{i96}. $r_e$ is the effective disk radius, 
 $M_{\rm B}$ is the absolute B-magnitude, and $h$ is the Hubble 
 constant $H_0$ defined as $h=H_0/100\,$km$\,$s$^{-1}$Mpc$^{-1}$.
}
\label{fig:mr-alpha2}
\end{figure*}

\begin{figure*}
   \epsfig{file=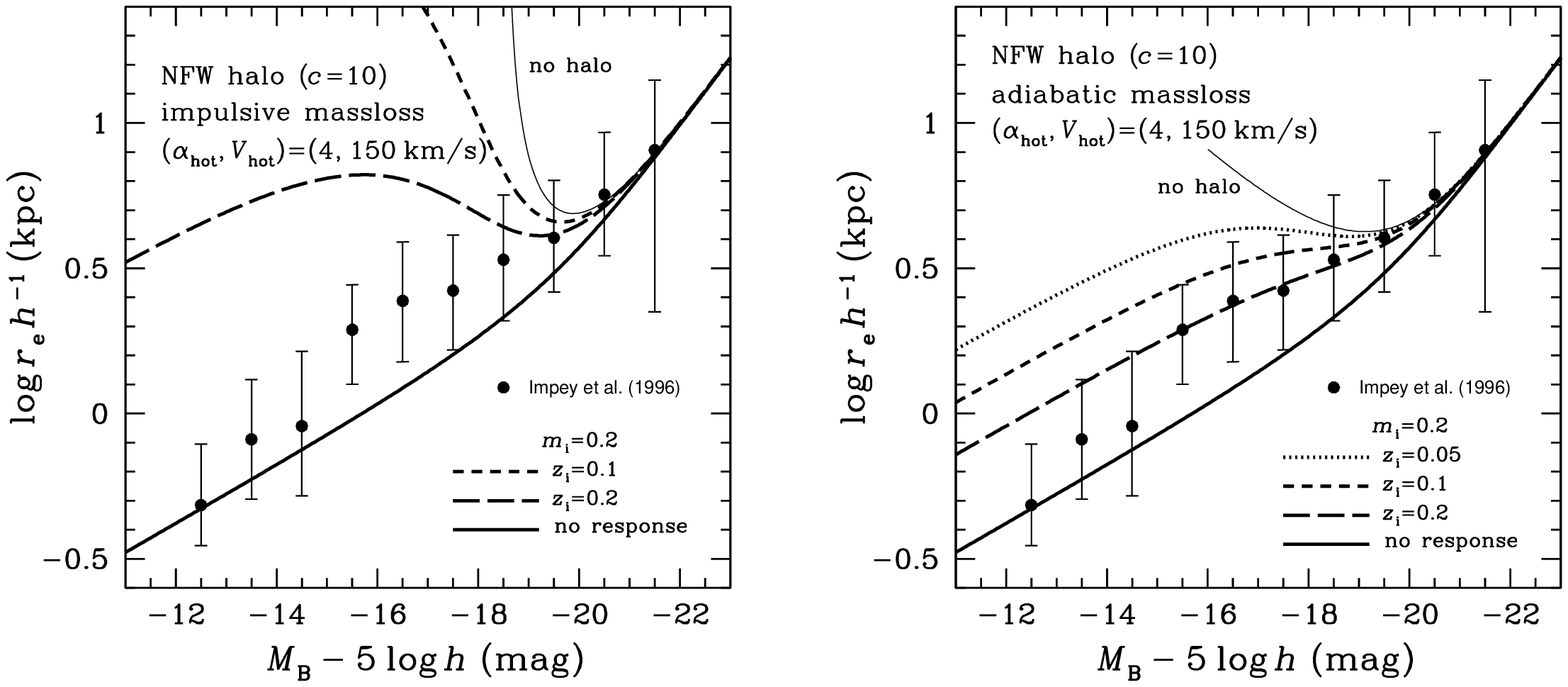,width=0.9\hsize}
 \caption{Theoretical B-band disc size versus absolute magnitude 
 relations with and without dynamical response of the Kuzmin disc 
 embedded in the NFW halo. Same as in Figure \ref{fig:mr-alpha2}, 
 but for $\alpha_{hot}=4$.
}
\label{fig:mr-alpha4}
\end{figure*}

This relation is then compared with the data of effective disc 
radius $r_e$ and absolute magnitude $M_{\rm B}$ of spiral galaxies in 
the B band taken from the table of \citet{i96}, and the results 
are shown in Figure \ref{fig:mr-alpha2} for $\alpha_{hot}=2$ and 
in Figure \ref{fig:mr-alpha4} for $\alpha_{hot}=4$, together with 
the data. In each of these figures the left and right panels are 
for the impulsive and adiabatic mass losses, respectively. 

In order to avoid any systematics arising from our use of the 
velocity data in the I band and the size data in the B band, 
we first transform the theoretical relation from the I band to the 
B band by applying the empirical formula 
$M_{\rm B}=0.86M_{\rm I}-1.31$, 
then we compare it with the data of $r_e$ and $M_{\rm B}$ in the B 
band. Such a transformation formula is obtained by eliminating 
$V_{rot}$ from the observed TFRs in the B and I bands \citep{pt88}, 
and gives B$-$I$\simeq 1.2$ for dwarf galaxies with $M_{\rm I}\simeq 
-18+5\log h$, and B$-$I$\simeq 1.6$ for bright galaxies with 
$M_{\rm I}\simeq -21+5\log h$. These colors agree 
well with the observed colors $\langle$B$-$V$\rangle=0.4-0.5$ 
and $\langle$V$-$I$\rangle=0.7-0.8$ for dwarf galaxies 
\citep{s95,s97,m99,o00}, and with the synthetic colors of 
$\langle$B$-$V$\rangle=0.5-0.6$ and $\langle$V$-$I$\rangle=1.0-1.1$ 
for bright Sb and Sc galaxies from Table 3a of \citet{f95}.

Here, applying the above transformation formula and allowing a 
small, constant offset of $\log r_e($B$)-\log r_e($I$)$ between 
the effective radii in the B and I bands, we have set the power 
index $\epsilon=2$ in equation (\ref{eq:sn-mimf2}) by adjusting
the $r_f-M_f$ relation along the horizontal and vertical axes in 
Figures \ref{fig:mr-alpha2} and \ref{fig:mr-alpha4} to fit to the 
observed disc size versus magnitude relation in the bright end. 
Our setting of $\epsilon=2$ is consistent with the scaling relation 
of dark haloes predicted by a power-law spectrum of initial density 
fluctuations with an index $n=-2$ which is a reasonable approximation 
to the CDM spectrum on galaxy scales \citep{nfw}.

We find that the $r_f-M_f$ relation is much more sensitive to the 
parameters than the $V_f-M_f$ relation, and many of the predicted 
relations cannot reproduce the faint data at all. In particular, 
all the relations considered for the impulsive mass loss should be 
rejected by the data. On the other hand, for the adiabatic mass loss, 
the relations of $z_i=0.1-0.2$ and $\alpha_{hot}=2$ agree with the 
data over the full range of $M_{\rm B}$ observed, and the relation of 
$z_i=0.2$ and $\alpha_{hot}=4$ agrees with the data brighter than  
$M_{\rm B}=-16$ only. 

\section{Results}
Current SA models, where the mass loss from individual galaxies 
by SN feedback is taken into account without considering dynamical 
response, are known to overpredict the rotation velocity of dwarf 
spiral galaxies (thick solid lines in Figures \ref{fig:tfr-nfw-alpha2} 
and \ref{fig:tfr-nfw-alpha4}) and at the same time underpredict 
their disc size (thick solid lines in Figures \ref{fig:mr-alpha2} 
and \ref{fig:mr-alpha4}). We could easily imagine that such 
discrepancies are resolved by the dynamical response of a virialized 
system associated with weakening of the gravitational potential, 
most likely owing to the mass loss that we consider in this paper. 

In fact, there exist several solutions with dynamical response 
that account for the observed relation between rotation 
velocity and absolute I-magnitude (Subsection 4.1) and the 
observed relation between disc size and absolute B-magnitude 
(Subsection 4.2). However, these two relations can only be 
reproduced simultaneously in a limited range of parameter 
space for the adiabatic mass loss, namely $m_i=0.2$ and 
$z_i=0.1-0.2$. For the SN feedback, the weak case of 
$\alpha_{\rm hot}=2$ is preferable to the strong case 
of $\alpha_{\rm hot}=4$.

Such an almost unique solution with $\alpha_{\rm hot}=2$ for 
the adiabatic mass loss gives the final ratio of $z_f=0.3-0.4$ 
for dwarf galaxies.  This value of $z_f$ agrees with observations 
in the local universe (see Section. \ref{sec:response}). 
The value of $(J/M)_f/(J/M)_i$ beyond unity suggests that the 
mass loss does not accompany the loss of angular momentum.

\begin{figure*}
   \epsfig{file=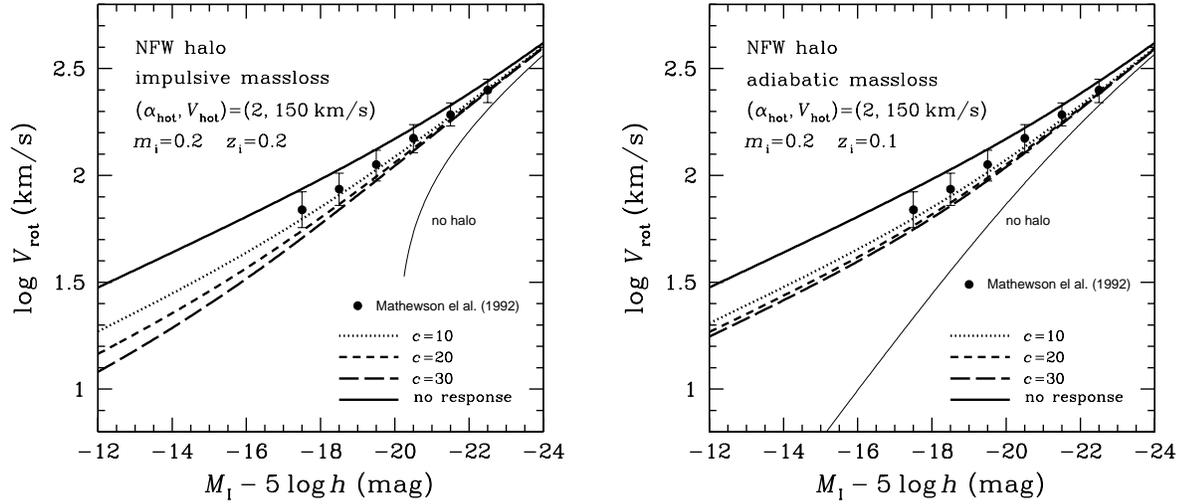,width=0.9\hsize}
 \caption{
Theoretical I-band Tully-Fisher relations with and without dynamical 
response of the Kuzmin disc embedded in the NFW halo.
Same as in Figure \ref{fig:tfr-nfw-alpha2}, but for the dependence on 
the concentration parameter $c$.  The results are shown by various 
thick lines in the left panel for the impulsive mass loss and in the 
right panel for the adiabatic mass loss.}
\label{fig:tfr-alpha2-c}
\end{figure*}

\begin{figure*}
   \epsfig{file=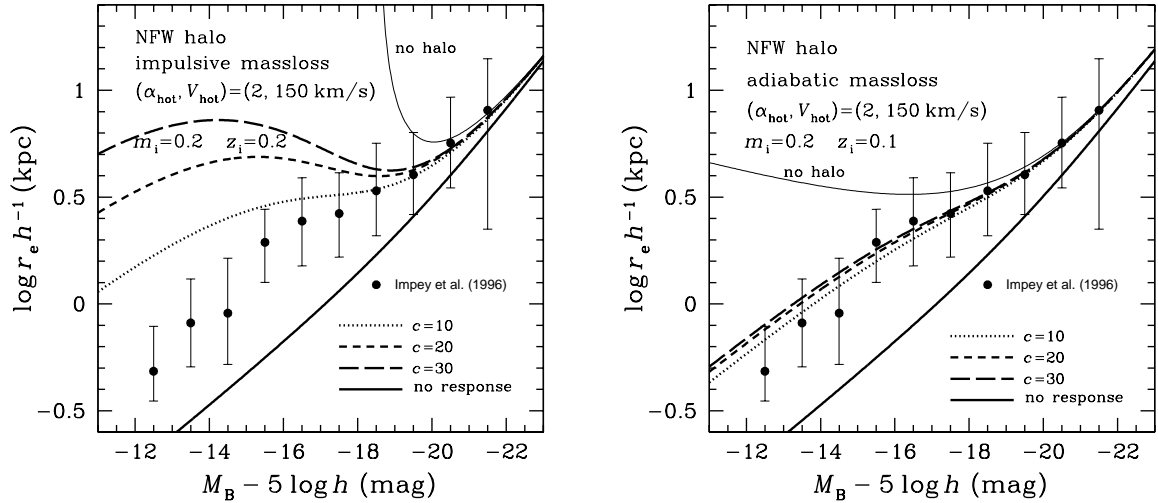,width=0.9\hsize}
 \caption{
Theoretical B-band disc size versus absolute magnitude relations 
with and without dynamical response of the Kuzmin disc embedded 
in the NFW halo.
Same as in Figure \ref{fig:mr-alpha2}, but for the dependence on 
the concentration parameter $c$. The results are shown by various 
thick lines in the left panel for the impulsive mass loss and in the 
right panel for the adiabatic mass loss.}
\label{fig:mr-alpha2-c}
\end{figure*}

The results above are obtained for dark haloes with a concentration 
parameter of $c=10$, but it is known that $c$ is predicted to vary 
from 10 for large haloes to significantly larger values of $20-30$ 
for small haloes \citep[]{b01b, m07}.  It is then worth examining 
this dependence with other parameters fixed.  The results by 
varying $c$ for the Tully-Fisher relation and the disk size versus 
magnitude relation are shown in Figures \ref{fig:tfr-alpha2-c} and 
\ref{fig:mr-alpha2-c}, respectively.  As seen from these figures, 
contrary to the case of impulsive mass loss, the variation in $c$ 
does not make much difference in the results for the adiabatic 
mass loss, which validates our conclusion in favor of adiabatic 
mass loss, irrespective of concentration of the mass included in 
dark haloes.

\section{Discussion}
\subsection{Losses of mass and angular momentum}
The angular momentum distribution in dark haloes is well 
established theoretically based on the CDM model 
\citep[e.g.][]{ct96, ng98}.  It is natural to assume that the 
baryonic components acquire the same amount of specific 
angular momentum as that of their host haloes, because the 
cosmic tidal field provides the angular momentum to dark matter 
and baryons in the same manner \citep{w84}. This assumption has 
been a basis of most theories of galaxy formation. 

From observations of dwarf spiral galaxies, \citet{vdb01a}
found that their mean (or {\it total}) specific angular momenta 
are very similar to those of their host haloes predicted by those 
theories. Moreover they found that the the disc to halo mass 
ratio is smaller than the baryon fraction in the Universe.  This 
means that only a part of baryons forms discs whereas it keeps an 
amount of specific angular momentum similar to that of host haloes. 
They also examined the specific angular momentum distribution 
in individual haloes, and surprisingly found that it is inconsistent 
with theoretical prediction by \citet{b01} based on high-resolution 
$N$-body simulations. 

Secondly, the observed surface density distribution of dwarf 
spiral galaxies is described as an exponential distribution 
\citep[e.g.][]{mfb92}, different from that expected from the CDM 
model which provides much more centrally concentrated discs when 
compared with the exponential distribution.  This cannot be solved 
even if SN feedback is simply considered \citep{vdb01b}.  

One possible solution to the above discrepancies should be obtained 
by considering the mass loss from the central region of discs, where 
the gas should have the specific angular momentum lower than that 
in the outer region.  SN feedback driving the mass loss without the 
loss of angular momentum leads discs to increasing their specific 
angular momentum, i.e.,  $(J/M)_{f}/(J/M)_{i}>1$.  This situation is 
consistent with that considered in this paper, as shown in the 
previous section.

So far we have assumed rotation-dominated discs throughout gas 
removal. With this assumption, the specific angular momentum always 
increases after the mass loss, yielding $(J/M)_{f}/(J/M)_{i}>1$.  
On the other hand, the transformation of rotation-dominated discs 
to velocity dispersion-dominated discs, which is the puff-up 
transformation, leads discs to decreasing the specific angular 
momentum, yielding $(J/M)_{f}/(J/M)_{i}<1$ \citep{bs79}.  
This situation should occur when the gas in the outer region is 
removed where the specific angular momentum is large.  Such a 
way of gas removal would be caused by the tidal stripping in clusters 
of galaxies \citep[e.g.][]{gg72,fn99,on03}.  \citet{bs79} proposed 
that this mechanism should be of lenticular galaxy formation.  
If so, the fraction of S0 galaxies in clusters will be higher than 
in fields, and the rotation velocity of dwarf galaxies in clusters 
will be systematically lower than that in fields.  This should be 
observationally clarified in future.

\subsection{Density profiles of dark haloes}
We have assumed three types of density profiles of dark haloes, 
such as the NFW profile with an inner slope of -1 and an outer 
slope of -3 (equation \ref{eq:density-nfw}), and two power-law 
profiles with a slope of 0 (homogeneous) and -1 ($r^{-1}$).  

The NFW profile is suggested by $N$-body simulations, and many 
authors claim that the inner slope will become even steeper 
after the condensation of baryons owing to the cooling 
\citep{bffp86}, which is called as the adiabatic contraction
(but see \citet{sm05} for the effects of random motions).  
In fact, this implicitly assumes that the cooling timescale is 
longer than the relaxation timescale for dark haloes to settle 
into the NFW profile, that is, baryons cool and shrink after 
virialisation.
On galaxy scales, 
however, this would not be the case. The cooling timescale is 
much shorter than the dynamical timescale, and the relaxation 
timescale is similar to or longer than the dynamical timescale.  
Thus it is not unnatural to assume that a mixture of dark matter 
and cooled baryons virializes.  In this case, there would be galaxy 
discs within dark haloes with the NFW profile, without undergoing 
the adiabatic contraction.  This might be an opposite limiting 
case to the adiabatic contraction, but recent X-ray observation 
of the intracluster medium has found that there is no evidence 
of contraction of the dark halo \citep{z06}.  Since the central 
cD galaxy is massive enough to suppress the SN feedback, we do 
not need to expect the expansion due to dynamical response to 
SN-induced gas removal.  Thus it is possible to say that a 
simple assumption of adiabatic contraction does not work in 
reality.  This should justify our use of the NFW profile for 
dark haloes that surround galaxy discs.  To clarify this, of 
course, high-resolution hydrodynamical simulations including 
the gas cooling processes are required \citep[e.g.][]{gkkn04}.

\subsection{Hierarchical formation of galaxies}
In Section 4, our results of dynamical response to SN-induced 
gas removal within dark haloes are applied to scaling relations 
for galaxies, under the assumption that simple scalings among 
mass, velocity, and size of baryonic components have been set 
up {\it before} the gas removal. It is reasonable to consider 
that these are inferred from observations of scaling relations 
for massive galaxies where SN feedback is not effective owing 
to their deep gravitational potential wells.

This seems to be somewhat in contrast to the approach of, for 
example, \citet{vdb01b}, in which the direction of angular 
momentum vector is assumed to be invariant.  However, it has 
been shown that the direction can be moved by contiguous accretion 
of dark matter \citep{ng98, sw04}. Therefore, it is not assured 
that the specific angular momentum of discs is similar to that 
of host haloes averaged over whole regions. 

We thus believe that it is reasonable to use the scaling relations 
for massive galaxies as the initial state for the dynamical response 
to gas removal, 

\section{Summary}
We have analyzed the dynamical response to SN-induced gas removal 
on rotation velocity and disc size of spiral galaxies, explicitly 
taking into account the underlying gravitational potential wells 
made by dark matter.  This is an extension of \citet{ny03} in which 
spherical galaxies were considered, but for thin discs in spherical 
dark haloes to describe spiral galaxies realistically.  Similarly, 
expansion of discs decreases their rotation velocity and increases 
their size owing to the dynamical response to gas removal even within 
dark haloes.  As shown in Section 3, the dynamical response provides 
unavoidable effects on the evolution of spiral galaxies, particularly 
on dwarf spiral galaxies because of their shallow gravitational 
potential wells. 

We have examined the effects of dynamical response on scaling 
relations among mass, rotation velocity, and disc size of spiral 
galaxies.  Since it is complicated to make such relations within a 
framework of hierarchical formation scenario of galaxies as discussed 
in Section 6, we use scaling relations which massive galaxies satisfy 
as the initial state before the gas removal. The SN-induced gas 
removal decreases the mass particularly of dwarf galaxies, thus 
distorting their scaling relations significantly.  Such a distortion 
in the TFR as well as the disc size versus magnitude relation has 
been pointed out by many papers working on SA models of galaxy 
formation \citep[e.g.][]{c00, ny04, nyeyg05}, in which the dynamical 
response on spiral galaxies is not taken into account.  We have shown 
that the dynamical response considered in this paper is able to 
reproduce the scaling relations as observed.

The hierarchical clustering scenario provides complicated galaxy 
formation processes. There are many physical processes such as 
radiative gas cooling, star formation and SN feedback, within merging 
dark haloes.  Furthermore, while the gas removal occurs mostly 
due to SN feedback in massive galaxies, other processes including 
photoheating by the UV background may unlikely scale the same 
way in dwarf galaxies \citep{ngs99, s02, bfbcl03, hygs06}.  In addition, the 
formation history is also complicated. For example, discs accrete     
the gas which is once expelled from galaxies.  It means that discs 
get bigger even after the dynamical response has exerted upon 
them.  Then, luminous discs tend to cause large extinction by dust, 
which would bend the TFR.  To investigate the observed properties 
of galaxies, therefore, we need to fully incorporate the effects of 
dynamical response into realistic models of galaxy formation like 
our SA models.  We will study the physical origin of scaling relations 
including the TFR in more realistic situations in a forthcoming paper.

\section*{ACKNOWLEDGMENTS}    
This work was supported in part by the Grant-in-Aid for the Scientific
Research Fund (17104002 and 18749007) of the Ministry of Education,
Culture, Sports, Science and Technology of Japan, and by a Nagasaki
University president's Fund grant.
HK is supported by the Japan Society for the Promotion of Science for
Young Scientists (1589).

\onecolumn
\appendix

\section{gravitational potential energy of two-component system}
In this section we provide analytic expressions of $f(z)$ and $q(z)$
for the gravitational interaction potential energy of the Kuzmin disk
of baryons embedded in various density distributions of dark matter halo.

\subsection{The NFW dark halo}
The NFW density distribution and the corresponding gravitational potential
are given respectively by
\begin{eqnarray}
 \rho(r)=\rho_d
c^3\left[\frac{cr}{r_d}\left(1+\frac{cr}{r_d}\right)^2\right]^{-1} ,
\end{eqnarray}
and
\begin{eqnarray}
 \Phi(r)=-4\pi G\frac{\rho_dr_d^3}{r}\ln\left(1+\frac{cr}{r_d}\right) .
\end{eqnarray}
Then, we obtain the total mass of dark halo
\begin{eqnarray}
 M_d=4\pi \rho_dr_d^3\left[\ln\left(1+c\right)-\frac{c}{1+c}\right] .
\end{eqnarray}
and the analytic expression of the following function:
\begin{eqnarray}
 f(z)=
c\left[\ln\left(1+c\right)-\frac{c}{1+c}\right]^{-1}
\Bigg[\frac{1}{cz}\ln \frac{cz}{2}
+\frac{1}{cz\sqrt{c^2z^2+1}}
\ln \frac{(\sqrt{c^2z^2+1}+1)(\sqrt{c^2z^2+1}+cz)}{cz}
\Bigg] .
\end{eqnarray}
For this case, we cannot obtain the exact analytic expression of
$q(z)$.  Instead, we obtain an approximate expression of $q(z)$
around $z=0$ as follows:
\begin{eqnarray}
 q(z)\simeq
c\left[\ln\left(1+c\right)-\frac{c}{1+c}\right]^{-1}
\left[
cz^2\left(\frac 7 9 + \frac 2 3 \log\frac{cz}{2}\right)
-\frac 43 c^2z^3
\right] .
\end{eqnarray}

\subsection{The homogeneous dark halo}
The homogeneous density distribution and the corresponding gravitational
potential are given respectively by
\begin{eqnarray}
 \rho(r)=\rho_d\theta(r_d-r) ,
\end{eqnarray}
and
\begin{eqnarray}
 \Phi(r)=
-4\pi G \rho_d
\left[-\frac{r^2}{6}+\frac{r_d^2}{2}+
\left(\frac{r^2}{6}-\frac{r_d^2}{2}+\frac{r_d^3}{3r}\right)\theta(r-r_d)\right]
.
\end{eqnarray}
Then, we obtain the total mass of dark halo
\begin{eqnarray}
 M_{d}=\frac{4}{3}\pi r_d^3 \rho_d ,
\end{eqnarray}
and the analytic expressions of the following functions:
\begin{eqnarray}
 f(z)=z^2+\frac{3}{2}+\frac{1}{z}-\frac{1}{z}(z^2+1)^{3/2} ,
\end{eqnarray}
and
\begin{eqnarray}
q(z)=-\sqrt{z^2+1}(2z^2-1)+2(z^3-2)+\frac{3}{z}\sinh^{-1}(z) .
\end{eqnarray}

\subsection{The $r^{-1}$ profile dark halo}
The $1/r$ density distribution and the corresponding gravitational potential
are given respectively by
\begin{eqnarray}
 \rho(r)=\rho_d \frac{r_d}{r}\theta(r_d-r) .
\end{eqnarray}
and
\begin{eqnarray}
 \Phi(r)=
-4\pi G \rho_d r_d^2
\left[-\frac{r}{2r_d}+1+
\left(\frac{r}{2r_d}-1+\frac{r_d}{2r}\right)\theta(r-r_d)\right] .
\end{eqnarray}
Then, we obtain the total mass of dark halo
\begin{eqnarray}
 M_{d}=2\pi r_d^3 \rho_d ,
\end{eqnarray}
and the analytic expressions of the following functions:
\begin{eqnarray}
 f(z)=2+z\ln z-z\ln(1+\sqrt{z^2+1})+\frac{1}{z}-\frac{\sqrt{z^2+1}}{z} ,
\end{eqnarray}
and
\begin{eqnarray}
 q(z)=
-4+\frac{4}{3}\sqrt{z^2+1}+\frac{8}{3z}\sinh ^{-1}(z)+\frac{4}{3}z^2\ln z
-\frac{4}{3}z^2\ln(1+\sqrt{z^2+1}) .
\end{eqnarray}

\bsp

\begin{thebibliography}{}
\bibitem[Benson et al. (2003)]{bfbcl03}
Benson A. J., Frenk C. S., Baugh C. M., Cole S.,  Lacey C. G., 2003,
MNRAS, 343, 679

\bibitem[Biermann \& Shapiro(1979)]{bs79}Biermann P., Shapiro S. L., 1979, ApJ, 230L, 33
\bibitem[Blumenthal et al.(1984)]{bfpr84}Blumenthal G. R., Faber S. M., Primack J. R., Rees M. J., 1984, Nature, 311, 517 
\bibitem[Blumenthal et al.(1986)]{bffp86}Blumenthal G. R., Faber S. M., Flores
				   R., Primack J. R., 1986, ApJ, 301, 27
\bibitem[Bullock et al.(2001a)]{b01}Bullock J. S., Dekel A., Kolatt
				   T. S., Kravtsov A. V., Klypin A. A.,
				   Porciani C., Primack J. R., 2001,
				   MNRAS, 321, 559
\bibitem[Bullock et al.(2001b)]{b01b}
Bullock J.S.,  Kolatt, T.S., Sigad, Y.,
Somerville, R.S., Kravtsov, A.V., Klypin, A.A., Primack, J.R.,
				   Dekel, A., 2001,
				   ApJ, 555, 240
\bibitem[Burkert(1995)]{bu95}Burkert, A., 1995, ApJL, v.447, 25
\bibitem[Catelan \& Theuns(1996)]{ct96}Catelan P., Theuns T., 1996,
				   MNRAS, 282, 436
\bibitem[Cole et al.(2000)]{c00}Cole S., Lacey C. G., Baugh C. M., Frenk C. S., 2000, MNRAS, 319, 168
\bibitem[Croton et al.(2006)]{croton06}Croton D. J., Springel V., White S. D. M. De Lucia G., Frenk C. S., Gao L., Jenkins A., Kauffmann G., Navarro J. F., Yoshida N., 2006, MNRAS, 365, 11
\bibitem[Dekel \& Silk(1986)]{ds86}Dekel A., Silk J., 1986, ApJ, 303, 39
\bibitem[De Lucia, Kauffmann \& White(2004)]{dlkw04}De Lucia G.,
				  Kauffmann G., White S. D. M., 2004,
				  MNRAS, 349, 1101
\bibitem[Faber(1982)]{f82}Faber S. M., 1982, in  Proceedings of the Study Week on Cosmology and Fundamental Physics, eds. H. A. Brueck, G. V. Coyne, \& M. S. Longair (Vatican City State, Pontificia Academia Scientiarum), p.191
\bibitem[Faber \& Gallagher(1979)]{fg79}Faber S. M., Gallagher J. S., 1979, ARAA, 17, 135
\bibitem[Faber \& Jackson(1976)]{fj76}Faber S.M., Jackson R. E., 1976, ApJ, 204, 668
\bibitem[Fujita \& Nagashima(1999)]{fn99}Fujita Y., Nagashima M., 1999,
				   ApJ, 516, 619
\bibitem[Fukugita et al.(1995)]{f95}
Fukugita, M., Shimasaku, K., Ichikawa, T., 1995, PASP, 107, 945

\bibitem[Gnedin et al.(2004)]{gkkn04}
Gnedin O. Y., Kravtsov A. V., Klypin A. A., Nagai D., 2004, ApJ, 616, 16

\bibitem[Governato et al.(2007)]{g07}
Governato F., Willman B., Mayer L., Brooks A., Stinson G.,
			      Valenzuela O., Wadsley J., Quinn T., 2007,
			      MNRAS, 374, 1479

			     
\bibitem[Gunn \& Gott(1972)]{gg72}Gunn J. E., Gott J. R., 1972, ApJ,
				   176, 1
\bibitem[Hoeft et al.(2006)]{hygs06}
Hoeft M.,  Yepes  G.,  Gottl\"ober S.,  Springel  V.,
2006, MNRAS, 371, 401
\bibitem[Impey et al.(1996)]{i96}
Impey, C. D., Sprayberry, D., Irwin, M. J., Bothun, G. D.,
  Mathewson, D. S., Ford, V. L., Buchhorn, M., 1996, ApJS, 105, 209
\bibitem[Kang et al.(2005)]{kjmb05}Kang X., Jing Y. P., Mo H. J., Borner G., 2005, ApJ, 631, 21
\bibitem[Kannappan, Fabricant \& Franx(2002)]{kff02}Kannappan S. J., Fabricant D. G., Franx M., 2002, AJ, 123, 2358
\bibitem[Kuzmin(1952)]{k52}Kuzmin G., 1952, Publ.Astr.Obs.Tartu, 32, 211
\bibitem[Kuzmin(1956)]{k56}Kuzmin G., 1956, Astron. Zh., 33, 27
\bibitem[Macci\'o et al.(2007)]{m07}
Macci\'o, A.V., Dutton, A.A., van den Bosch, F.C., Moore, B.P., 
  Doug, S., 2007, MNRAS, 378, 55
\bibitem[Makarova(1999)]{m99}Makarova, L, 1999, A\& AS, 139, 491
\bibitem[Mathewson, Ford \& Buchhorn(1992)]{mfb92}Mathewson D. S., Ford V. L., Buchhorn M., 1992, ApJS, 81, 413
\bibitem[Nagashima \& Gouda(1998)]{ng98}Nagashima M., Gouda N., 1998,
				   MNRAS, 301, 849
\bibitem[Nagashima et al.(1999)]{ngs99}
Nagashima M., Gouda N.,Sugiura N., 1999, MNRAS, 305, 449
\bibitem[Nagashima \& Yoshii(2003)]{ny03}Nagashima M., Yoshii Y., 2003,MNRAS, 340, 509
\bibitem[Nagashima \& Yoshii(2004)]{ny04}Nagashima M., Yoshii Y.,2004, ApJ, 610, 23
\bibitem[Nagashima et al.(2005)]{nyeyg05} Nagashima M., Yahagi H., Enoki M., Yoshii Y \& Gouda N. 2005, ApJ 634, 26
\bibitem[Navarro, Frenk \& White(1997)]{nfw}Navarro J.F., Frenk C.S., White S.D.M., 1997, ApJ, 490, 493
\bibitem[Okamoto \& Nagashima(2003)]{on03}Okamoto T., Nagashima M.,
				   2003, ApJ, 587, 500
\bibitem[O'Neil et al.(2000)]{o00}
O'Neil, K., Bothun, G. D.,  Schombert, J., 2000, AJ, 119, 136
\bibitem[Pierce \& Tully(1988)]{pt88}Pierce M. J., Tully R. B., 1988, ApJ, 330, 579
\bibitem[Pierini(1999)]{p99}Pierini D., 1999, A\&A, 352, 49
\bibitem[Pildis, Schombert \& Edger(1997)]{pse}Pildis, R. A., Schombert,
				 J. M., Eder, J. A., 1997, ApJ, v.481,
				 157
\bibitem[Portinari \& Sommer-Larsen(2007)]{psl07}Portinari L.,
				   Sommer-Larsen J., 2007, MNRAS, 375, 913
\bibitem[Saitoh \& Wada(2004)]{sw04}Saitoh T. R., Wada K., 2004, ApJL,
				   615, L93
\bibitem[Sakai et al.(2000)]{s00}Sakai S. et al., 2000, ApJ, 529, 698
\bibitem[Schombert et al.(1995)]{s95}Schombert J. M., Pildis R. A., Eder J. A., Oemler A., 1995, AJ, 110, 2067
\bibitem[Schombert et al.(1997)]{s97}Schombert, J. M., Pildis, R. A.,
			      Eder, J. A., 1997, ApJS, v.111, 233
\bibitem[Sellwood \& McGaugh(2005)]{sm05}
Sellwood J. A., McGaugh S. S., 2005, ApJ, 634, 70

\bibitem[Somerville \& Primack(1999)]{sp99}Somerville R.S., Primack
				   J. R., 1999, MNRAS, 310, 1087
\bibitem[Sommerville(2002)]{s02}Somerville R.S., 2002, ApJ, 572, L23


\bibitem[Steinmetz \& Navarro(1999)]{sn99}Steinmetz, M., Navarro J. F., 1999, ApJ, 513, 555
\bibitem[Tully \& Fisher(1977)]{tf77}Tully R. B., Fisher J. R., 1977, A\&A, 54, 661
\bibitem[van den Bosch et al.(2001a)]{vdb01a}van den Bosch F. C., Burkert A.,
				   Swaters R. A., 2001a, MNRAS, 326, 1205
\bibitem[van den Bosch(2001b)]{vdb01b}van den Bosch F. C., 2001b, MNRAS, 327, 1334
\bibitem[van den Bosch(2002)]{vdb02}van den Bosch F. C.,2002, MNRAS, 332, 456
\bibitem[White(1984)]{w84}White S.D.M., 1984, ApJ, 286, 38
\bibitem[Willick et al.(1997)]{w97}Willick J. A., Courteau S., Faber S. M., Burstein D., Dekel A., Strauss M. A., 1997, ApJS, 109, 333 
\bibitem[Yoshii \& Arimoto(1987)]{ya87}Yoshii Y., Arimoto N., 1987,A\&A,
				  188, 13
\bibitem[Zappacosta et al.(2006)]{z06}Zappacosta L., Buote D. A.,
				   Gastaldello F., Humphrey P. J.,
				   Bullock J., Brighenti F., Mathews W.,
				   2006, ApJ, 650, 777
\end{thebibliography}
\end{document}